\documentclass[manuscript]{acmart}

\AtBeginDocument{%
  }

\setcopyright{acmlicensed}
\copyrightyear{2026}
\acmYear{2026}
\acmDOI{XXXXXXX.XXXXXXX}

\acmConference[]{}{2026}{Woodstock, NY}
\acmISBN{978-1-4503-XXXX-X/18/06}

\usepackage{tabulary}
\usepackage[true]{anonymous-acm}
\usepackage{subcaption}

\usepackage[colorinlistoftodos]{todonotes}

\begin{document}

\title{Friction in AI-Assisted Clinical Decision-Making: A Case Study on The Role of Questions and `What-if' Scenarios}

\author{Simon W.S. Fischer}
\email{simon.fischer@donders.ru.nl}
\orcid{0000-0003-2992-6563}
\affiliation{%
  \institution{Donders Institute for Brain, Cognition, and Behaviour, Radboud University}
  \department{Dpt. of Human-Centred Intelligent Systems}
  \city{Nijmegen}
  \country{The Netherlands}
}

\author{Hanna Schraffenberger}
\email{hanna.schraffenberger@ru.nl}
\orcid{0000-0003-1847-2754}
\affiliation{%
  \institution{Interdisciplinary Hub for Digitalization and Society (iHub) and Institute for Computing and Information Sciences (iCIS), Radboud University}
  \city{Nijmegen}
  \country{The Netherlands}
}
\author{Miranda L. van Hooff}
\orcid{0000-0001-5313-6436}
\affiliation{%
  \institution{Radboud UMC}
  \department{Dpt. of Orthopaedic Surgery}
  \city{Nijmegen}
  \country{The Netherlands}
}
\affiliation{%
  \institution{Sint Maartenskliniek}
  \city{Nijmegen}
  \country{The Netherlands}
}

\author{Serge Thill}
\email{serge.thill@donders.ru.nl}
\orcid{0000-0003-1177-4119}
\affiliation{%
  \institution{Donders Institute for Brain, Cognition, and Behaviour, Radboud University}
  \department{Dpt. of Human-Centred Intelligent Systems}
  \city{Nijmegen}
  \country{The Netherlands}
}

\author{Pim Haselager}
\email{pim.haselager@donders.ru.nl}
\orcid{0000-0002-4077-9560}
\affiliation{%
  \institution{Donders Institute for Brain, Cognition, and Behaviour, Radboud University}
  \department{Dpt. of Human-Centred Intelligent Systems}
  \city{Nijmegen}
  \country{The Netherlands}
}

\renewcommand{\shortauthors}{Fischer et al.}

\begin{abstract} %
  A major challenge in AI-supported decision-making is achieving appropriate reliance on AI. To reduce overreliance on AI, approaches for cognitive friction have been proposed that promote reflection and critical thinking. In this paper, we investigate how two such approaches are perceived by medical experts in a real-world decision task. We replicated an existing decision support system used in clinical practice and extended it with outputting reflective questions and what-if scenarios. We gathered clinicians’ ($n=7$) feedback on this prototype through in-situ interviews. While the questions were perceived as unhelpful for reflective thinking in itself, their potential was seen in serving as reminders to consider relevant information, e.g., when attention wanes. Inspecting what-if scenarios was found useful for considering alternative outcomes and informing discussions with patients. Our participants regarded the prototype as a promising training tool for novice clinicians. From the clinicians’ feedback, we make recommendations for designing friction in human-AI interactions.
\end{abstract}


\received{\today}

\maketitle

\section{Introduction}
\label{sect:intro}

In healthcare, decision support systems (DSS) are used for diagnosis, treatment prediction, and resource allocation \cite{Lindvall2021, Sutton2020, Walsh2019, Zhang2024}. Clinical DSS have the potential to improve patient care by providing additional information, such as predictions or recommendations, to help physicians make decisions \cite{Ouanes2024}. 
At the same time, however, studies point to the risk that decision-makers, including experts, may accept incorrect outputs from DSS \cite{Klingbeil2024,Vasconcelos2023}. This is referred to as overreliance on DSS, which can lead to harmful consequences such as misdiagnoses \cite{Jacobs2021a, Dratsch2023}. %

To mitigate overreliance, so-called frictional design approaches or cognitive forcing functions have been proposed, which intentionally introduce interruptions into human-AI interactions. As such, some systems provide recommendations only once the decision-maker has made a (provisional) decision \cite{Fogliato2022}, or they do not provide any recommendation but instead provide explanations both for and against a decision \cite{Miller2023}. Similarly, it is suggested that the DSS should challenge the decision-maker \cite{Sarkar2024}, whether through so-called provocations \cite{Drosos2025} or through a `reflection machine' that raises questions \cite{Cornelissen2022}. These approaches for cognitive friction hold potential to reduce overreliance because they can promote the cognitive engagement of the decision-maker in the form of reflection and critical thinking \cite{Cox2016, Gould2021, Natali2025}.

Reflection helps bring subconscious considerations into conscious awareness \cite{Sengers2005}, and, as \citet{Dewey1933} notes, reflection is an inquiry into one's own understanding, assumptions, and gaps in thinking. Reflection thus has been shown to improve decision-making \cite{Mamede2008, Hess2015, Prakash2019}. Similarly, critical thinking is the process of analysing and evaluating information, as well as considering alternative courses of action \cite{Facione2015}. One way to stimulate reflection and critical thinking, i.e., cognitive engagement, is through questions \cite{Ishtiaq2024, Ho2023, Osmond2005}.
It therefore seems desirable for a DSS to not only provide recommendations or predictions but also to raise questions that help decision-makers reflect on the decision at hand and enable them to investigate alternative outcomes, for example through hypothetical questions or what-if scenarios.

The potential of questions in human-AI decision-making remains largely underexplored and nascent. Recent studies on promoting reflection through guided questions \cite{Li2026a} and counterfactual what-if questioning \cite{Tian2026} show promising effects on reducing overreliance on DSS and increasing decision accuracy. Similarly, two other studies found that questions can improve the assessment of information validity \cite{Danry2023} and curb impulsive decisions in a financial investment task \cite{Reicherts2022}. All of those studies, however, focus on either toy examples \cite{Li2026a,Tian2026}, are evaluated with laypeople or crowd workers who lack relevant expertise \cite{Li2026a,Danry2023,Tian2026}, or do not focus on a decision task involving a DSS that provides predictions \cite{Danry2023,Reicherts2022}.
It therefore remains uncertain how well questions, and friction approaches for promoting cognitive engagement in general, perform in real-world settings and within existing DSS-informed workflows \cite{Lai2023a,Raees2026,Eckhardt2026}.

To complement existing studies with insights from a real-world setting, we investigate how two such forms of friction, namely questions and what-if scenarios, fit into professional practice and whether they are perceived as valuable by experienced decision-makers. In particular, we focus on the medical context, where clinicians have been using a DSS in clinical practice for several years. In view of this, we have the following research question (RQ) and design objective (DO):
\begin{itemize}
    \item \textbf{RQ.} How do clinicians perceive two friction strategies, namely questions and what-if scenarios, for their decision-making in clinical practice?
    \item \textbf{DO.} Develop a proof-of-concept prototype that outputs questions based on input data (i.e., patient information) and DSS predictions, and also allows for the manual creation of what-if questions or scenarios.
\end{itemize}

We conducted in-situ interviews with clinical spine specialists ($n=7$) from the same hospital department in order to examine the potential of our proof-of-concept prototype. %
For our design objective (DO), we replicated and extended a DSS used by a total of 11 clinicians (including our 7 participants) in their day-to-day decision-making regarding the treatment of chronic low back pain. 
Compared to the original DSS, our prototype not only displays the predictions regarding treatment outcomes, but also generates questions, such as \textit{``Is it possible to change the smoking behaviour? This would increase the predicted effectiveness of surgery by 20\%''} (Table~\ref{table:questions}).
In order to manually create hypothetical what-if questions or scenarios, we implemented an interactive element (Fig.~\ref{fig:sliders}). Using sliders, clinicians can modify the input data (i.e., patient information), and the corresponding DSS predictions are updated and displayed accordingly. The interactive visualisation thus offers insight into the question, \textit{``What is the prediction, if this input changes?''}

This paper makes the following contributions: 1) We present a proof-of-concept prototype that outputs questions and enables the creation of hypothetical what-if scenarios. 2) We provide insights from a real-world, complex, and open-ended clinical decision task regarding how experienced decision-makers perceive these two forms of friction (i.e., questions and what-if scenarios). 3) Based on the clinicians' feedback, we make recommendations for the design of friction strategies in professional practice.

\section{Background and Related Work}
\label{sect:background}

We focus on human-AI decision-making in the workplace. Relevant literature thus comes from human-computer interaction, psychology, and computer-supported cooperative work.
We will discuss approaches to friction in human-AI interaction that aim to stimulate reflection and critical thinking, both in general and specifically through questions.

\subsection{Frictional Design in Human-AI Decision-Making}
\label{sect:background-friction}

Frictional design intentionally adds (cognitive) interruptions, or micro-boundaries that slow down the decision-making process \cite{Cox2016, Gould2021, Natali2025}. Friction aims to facilitate the transition from fast, automatic thinking to slower, more deliberate thinking \cite{Chen2024}. In human-AI decision-making, friction ideally helps to mitigate the potential negative effects of adopting incorrect DSS recommendations, i.e., overreliance \cite{Bucinca2021,deJong2025}.

Various design approaches have been suggested to promote the decision-maker's cognitive engagement, reflection, and critical thinking during human-AI decision-making. Some examples include: allowing the decision-maker to make a provisional decision before the AI advice is displayed \cite{Bucinca2021, Fogliato2022}, delaying the presentation of recommendations \cite{Park2019,Bucinca2021}, providing second opinions on AI recommendations \cite{Lu2024}, providing confidence scores of the prediction \cite{Zhang2020,Li2025}, providing partial explanations \cite{deJong2025}, or providing evidence both for and against a decision instead of a recommendation \cite{Cabitza2025, Miller2023}. 

One study, which, like ours, focuses on a real-world decision-making task and is evaluated with experts, was conducted by \citet{Bach2023}. They investigate whether three different strategies can help to mitigate anchoring bias (i.e., the decision being influenced by the information perceived first) in clinical decision-making. To this end, \citet{Bach2023} implemented their strategies into an existing decision support system that ophthalmologists use to detect diabetic retinopathy in patients. The three strategies are, 1) \textit{hear the story first}, in which the decision-maker has to make their decision before seeing the DSS advice \cite{Fogliato2022}, 2) \textit{decision justification}, which ``prompts decision-makers to explain their reasoning to activate reflective thinking'' \cite{Bach2023}, and 3) \textit{consider the opposite}, which prompts ``the decision-maker to consider information that contradicts their current beliefs'' \cite{Bach2023}. 
The study participants, six ophthalmologists, rated the `consider the opposite' strategy as most useful, yet expressed concerns about being fact-checked by the DSS. The authors therefore advise for more subtle design patterns during human-AI interaction. In addition, \citet{Bach2023} note that their mitigation strategies are limited to a specific use case, which is why they call for similar studies in different healthcare settings than ophthalmology.

Following this call, we examine how healthcare professionals experience two forms of friction, namely receiving questions and investigating what-if scenarios. 
Our approach aims to encourage reflection on the reasons for a decision and the consideration of alternatives to the DSS advice. It therefore comes close to the strategies `decision justification' (2) and `consider the opposite' (3) presented by \citet{Bach2023}. Moreover, questions offer a more subtle design pattern than the three mitigation strategies.
Based on a realistic and complex decision task, namely the treatment of chronic low back pain, we aim to illuminate the effect of friction strategies on the decision-making process and how they can be integrated into existing workflows.

\subsection{Stimulating Reflection Through Questions}

Stimulating reflection and critical thinking through AI-generated questions has been examined in various studies, including in the field of education \cite{Orynbassarova2024, Xi2026, Favero2025}, in writing tasks \cite{Hugenroth2026,Chen2026}, in data annotation \cite{Khadar2025}, for ethical inquiry \cite{Li2025a}, and in the detection of misinformation \cite{Lim2026,Danry2023}.
In the context of decision-making involving a DSS, however, only few have investigated the potential of questions \cite{Tian2026,Li2026a}.

\citet{Li2026a} propose a guided, three-step reflection process consisting of questions that help 1) to detect conflicts between the AI prediction and human judgement, 2) to evaluate both supporting and contradicting perspectives (i.e., ``cognitive decoupling''), and 3) to provide reasons for the final decision. In particular, these questions aim to encourage engagement with the system's recommendation and explanation, e.g., ``Which parts of the AI's suggestion and explanation are reasonable?'' The authors report an increase of decision accuracy and a reduction of overreliance on DSS as a result of their approach.

\citet{Tian2026} propose questions that encourage counterfactual perspective-taking. For this, the DSS adopts the human's perspective and performs counterfactual analysis by data perturbation, to then display what-if questions, such as ``If you considered feature x, how would that change your confidence in your prediction?'' This helps decision-makers to examine their reasoning and how it might change under hypothetical modifications, as \citet{Tian2026} note.
As such, counterfactual reasoning serves to provide decision-makers with actionable insights to demonstrate `what would have (had) to be different in order to achieve a desirable outcome' \cite{Karimi2021,Wachter2017}. Counterfactual reasoning can thus be particularly helpful in healthcare, where different interventions need to be weighed against one another \cite{Prosperi2020}. In our prototype, we use a similar counterfactual question type based on feature perturbation as adopted by \citet{Tian2026}. In addition, we allow for manual creation of what-if scenarios, similar to interactive explanations \cite{Sokol2020,Meyer2024}.
As \cite{Tian2026} found, the counterfactual questioning reduced overreliance on the DSS predictions. 

Another prominent study by \citet{Danry2023} shows that questions, compared to causal explanations, can improve the discernment of logical statements. For their study, they present participants with valid and invalid statements, and provide either 1) a causal explanation related to the statement, or 2) a question adapted from the causal explanation in the form of ``If \textit{premise} does it follow that \textit{claim}?''. The findings suggest that questions improve the assessment of these statements. A reason could be that questions promote and lead to self-explanations, which improve understanding \cite{Chi1994}.

For a financial investment task, \citet{Reicherts2022} present a chatbot that asks expert investors questions about their investment hypotheses and the reasoning behind them. %
For example, the bot asks whether recent news has any relevance for a decision (i.e., availability bias). Overall, the participants found those questions effective in preventing impulsive actions. Furthermore, in an interview as part of another study on investment decisions, one participant voiced the need for reflective questions, \textit{``Even if just a simple message appears asking if this price is truly appropriate, I think I would calm down and reconsider''} \cite{Nam2025}. %
 
All these studies report positive effects of questions \cite{Danry2023,Reicherts2022,Li2026a,Tian2026}. Nevertheless, only two of them focus on a decision task involving a DSS that provides recommendations or predictions \cite{Li2026a,Tian2026}. Moreover, despite \cite{Li2026a} focuses on a medical task (i.e., diabetic prediction), the participants are laypeople and their expertise in the task is not disclosed. In addition, as the authors acknowledge, the decision task is rather simple and consists of a ground truth. Similarly, the decision task in \cite{Tian2026} concerns house price predictions, and thus resembles a toy example, which is evaluated with laypeople.
In view of this, we investigate whether (counterfactual) questions can help stimulate reflection during a complex real-world decision-making task in which experienced decision-makers work with a DSS.

Since there are various types of questions, a question taxonomy for AI-supported decision-making has been proposed \cite{Fischer2025a}. The taxonomy, which is based on concepts of reflection and critical thinking, identifies ten elements in machine-assisted decision-making that questions can relate to. Accordingly, to name a few, questions can relate to the input data, the dataset used to train the DSS, the causal structure of the recommendation, and alternatives to the recommendation. %
We will use this question taxonomy to identify relevant questions that our prototype can present.

\section{Use Case and Decision Task: Treatment of Chronic Low Back Pain}
\label{sect:use-case}

To investigate whether questions and what-if scenarios can promote cognitive engagement among decision-makers in professional practice, we focus on a real-world use case from the medical field, namely the treatment of chronic low back pain (CLBP). CLBP is a complex condition, the exact cause of which is often unknown \cite{Coupe2016}. In many cases, it is thus uncertain whether a treatment will result in improvement. Hence, there is no clear `ground truth'. To more effectively treat patients with a wide range of available treatment options, two approaches can help, namely 1) classification systems that classify a heterogeneous patient group into subgroups that are more likely to benefit from a specific treatment, and 2) clinical prediction rules that estimate the likelihood of a treatment outcome based on meaningful patient characteristics \cite{Coupe2016}.

We collaborated with hospital clinicians from a spinal department who work with a DSS that combines both approaches. The DSS considers three treatment options: spine surgery, a combined physical and psychological program, and counselling by the surgeon with referral to physiotherapy. Furthermore, the DSS categorises patients into two outcome groups for each treatment: `responder' (success) and `non-responder' (failure), depending on whether the patient's functional status is expected to improve as a result of the treatment. 
Overall, this results in a total of six treatment prognoses (3 treatments $x$ 2 outcomes), which are calculated independently with slightly different variables (Fig~\ref{fig:prototype}).
The DSS makes predictions based on tabular data from a patient self-report questionnaire. The questionnaire covers five areas: socio-demographic, pain, psychological, somatic, and quality of life.

To make decisions about how to treat the patients' chronic low back pain, the clinicians use the DSS in two ways: first, for patient triage to ensure the patient sees the right specialist, and second, to prepare for and during the patient consultation.
The predictive factors, as well as the clinical prediction rules, were defined by researchers at the same hospital where the DSS is used in clinical practice. %
This means the DSS could be tailored to the hospital's patient population. Nevertheless, due to the complexity of CLBP, the provided predictions should be viewed critically. With our prototype that presents questions and what-if scenarios, we want to assess whether it can support clinicians with this critical and reflective decision-making process.

\subsection{Our Prototype}
\label{sect:prototype}

\begin{figure}
    \centering
    \includegraphics[width=0.9\linewidth]{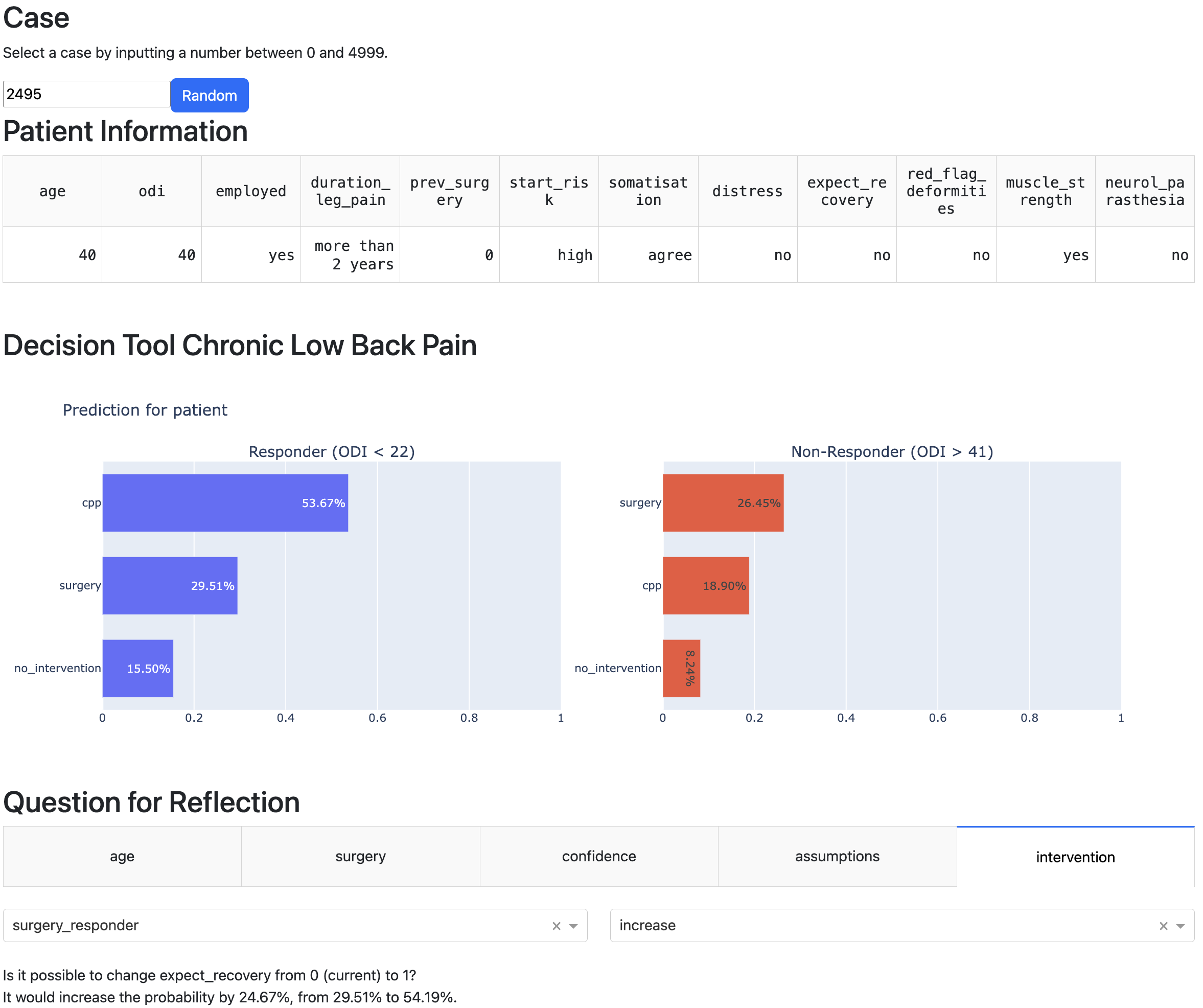}
    \caption{The interface of our prototype. It is possible to select a synthetic patient case, for which the information is displayed below. The predictions are shown for the three treatment options, divided into responder (success) and non-responder (failure) categories. At the bottom, questions are displayed in plain text. The current tab displays the seventh question, which computes perturbations for a given case to produce a forward-looking what-if question.}
    \label{fig:prototype}
    \Description{The image depicts an interface of a decision support system, showing the predictions of six treatment options in the form of bar charts. Below, questions are shown about the predictions.}
\end{figure}

With regard to our design objective (DO), we developed a proof-of-concept prototype in the form of an interactive Dash\footnote{\url{https://dash.plotly.com/}} app that outputs questions alongside treatment predictions (Fig.~\ref{fig:prototype}). To this end, we replicated and extended the DSS used at the spinal department ($DSS_{clinic}$).
A clinical researcher involved in its development, also an author of this paper, provided us with relevant information, enabling us to create a functional replica ($DSS_{replica}$) of the $DSS_{clinic}$. Moreover, during a preparatory meeting, a clinician gave us an introduction to how the $DSS_{clinic}$ is used in practice, and showed us how the predictions and the patient data, including the answers to the questionnaire, are presented. This allowed us to design our interface and the presentation of treatment predictions in a way that is familiar to the clinicians, namely in the form of bar charts divided into two categories: responder (success) and non-responder (failure), with one bar for each treatment option (Fig.~\ref{fig:prototype}).  %

\subsubsection{Dataset}

As we did not have access to real patient data, we generated tabular data using Mimesis\footnote{\url{https://mimesis.name/}}. Our data is based on the data schema of the patient questionnaire, which is also used for the $DSS_{clinic}$. The questionnaire consists of several indicators, each of which consists of sub-questions. For example, one variable in the patient questionnaire is the Oswestry Disability Index (ODI) score, which indicates the disability caused by low back pain. The score ranges from 0 to 100 and is calculated based on 10 sub-items, with higher scores indicating greater disability.
When predicting the treatment outcomes, the $DSS_{clinic}$ takes into account only the aggregated responses of the patient questionnaire. We therefore generated only the final scores for our purposes, not the responses to individual questionnaire items. 
So although our data set is more compressed than the actual questionnaire data, this has no impact on the functionality of our $DSS_{replica}$. For both $DSS_{clinic}$ \& $DSS_{replica}$, a prediction for a single patient is based on 12 features, consisting of binary (e.g., expected recovery), ordinal (e.g., pain levels), and continuous values (e.g., ODI score). Therefore, the schema of our synthetic data does not differ from real data. Moreover, we created a total of 5000 patient cases, so if necessary, clinicians could point out unrealistic cases during the interviews and we would then select another case.

\subsubsection{Question Generation}
\label{sect:question-generation}

To identify relevant questions that our prototype could output, we were inspired by four categories of the aforementioned question taxonomy \cite{Fischer2025a}, resulting in questions about 1) the case category, 2) assumptions and expectations of the decision-maker, 3) change intervention, and 4) model behaviour. In total, we have seven questions, with a balance of data-driven and general, non-data-driven questions (Table~\ref{table:questions}).
Three of the questions are based on the DSS behaviour and input data, namely those relating to the patient's age, previous surgeries, and hypothetical scenarios. The question about the clinician's confidence is in-between: it is rather general, yet contains data as point for reference.
The other three questions, all relating to the assumptions of the decision-maker, are more general and not data-driven.
In the prototype, the questions are presented below the predictions and in tabs, in order to display only one question at a time (Fig.~\ref{fig:prototype}).

\begin{table}[]
    \caption{Our prototype presents the following questions, alongside the treatment predictions. These questions build on a taxonomy of questions for critical reflection in machine-assisted decision-making \cite{Fischer2025a}.}
    \label{table:questions}
    \centering
    \begin{tabulary}{\linewidth}{@{}p{1.7cm}LL@{}}
    \toprule
         \textbf{Variable} &
         \textbf{Taxonomy Mapping} &
         \textbf{Question}
         \\
    \midrule 
         Age &
         Model Behaviour &
         (1) Is the patient's age of 47 years relevant to consider in this case? The patient of 47 years is approaching the red-flag threshold which is 50 years. Age correlates with the effect of surgery.
         \\
         Previous Surgery &
         Case Information &
         (2) When was the specified surgery performed and at which location of the spine? Previous surgeries reduce the effect of surgery by 15\% - 25\%.
         \\
         Confidence of Clinician &
         Assumptions and expectations of decision-maker &
         (3) How confident are you about your decision? The confidence of the prediction for the most effective treatment (surgery 59.92\%) is at 42.58\%.
         \\
         Assumptions of Clinician &
         Assumptions and expectations of decision-maker &
         (4) What are you taking for granted here?\newline
         (5) Does the prediction match your assumptions?\newline
         (6) Does the prediction change your initial judgement? If so, why?
         \\
         Hypothetical &
         Change Intervention &
         (7) Is it possible to change \textit{expected recovery} from `no' (current) to `yes'? It would increase the predicted effectiveness of surgery by 24.67\%, from 24.67\% to 54.19\%.
         \\
    \bottomrule
    \end{tabulary}
\end{table}

The first question concerns the functionality of the DSS, taking into account a built-in threshold. The DSS correlates the patient's age with the predicted effect of surgery. If the age crosses a certain threshold of 50 years, the prediction changes. In our prototype, a question is displayed when a patient approaches this threshold ($T\pm5$), asking whether the patient's age should receive more attention. We selected this question in order to reduce the static, fixed nature of a single threshold value.

The second question concerns the patient's information (i.e., case data). Previous operations influence the effectiveness of subsequent operations. It is important, however, to consider when and where on the spine previous surgeries were performed. If the previous surgery was performed 20 years ago, further surgeries are likely to be more effective compared than if the previous surgery was performed two years ago. %
Similarly, if the previous surgery was performed on a completely different part of the spine, its effect on further surgery is negligible. Hence, this question prompts the decision-maker to further contextualise certain information. 
In addition, we added an estimate of the influence of the factor `previous surgery' on the outcome, which is similar to a feature contribution. We use feature perturbations to calculate this influence, which we will describe in more detail below when discussing the seventh question.

The third question asks how confident the decision-maker is about their decision. To provide the clinician with a reference point and enable a comparison between their own assessment and the DSS prediction, we include the confidence score for the prediction with the highest score. To calculate the confidence value, the probability of the most effective treatment is normalised over all three treatment probabilities of one category, i.e., responder or non-responder.

The fourth, fifth, and sixth questions address the clinician's assumptions and whether the DSS predictions confirm or change those assumptions. For this, the prototype outputs general questions that are not based on data. The questions nevertheless require the clinicians to be more explicit about initial considerations regarding the case. %

The seventh question concerns hypotheticals and possible interventions to increase or decrease the likelihood of a desired outcome. Similar to counterfactual reasoning, this approach checks what the output (i.e., prediction) would be if the input (i.e., patient information) were different. 
Our prototype uses this information, however, to call into question the possibility of changing said value to make a desired treatment outcome more likely, e.g., \textit{``Is it possible to change factor \textit{x}? This would increase expected effectiveness by 20\%''}. So instead of a backward-looking evaluation, as is the case with counterfactuals, we provide a forward-looking (what-if or `how-to-be-that') hypothetical. To this end, we perturb the features of the patient case, compare the predictions of each perturbation to the original case, and search for the maximum change in outcome (prediction) with the minimum change in input (patient data). To specify the desired outcome, drop-down boxes allow to select the desired treatment and whether its expected result should increase or decrease relative to the current prediction (Fig.~\ref{fig:prototype}).

Although we have selected these seven questions, it is, of course, possible to generate more or different questions. In our case, the insights gained from the DSS functionality were the most promising and obvious approaches. Besides, counterfactual reasoning and the provision of confidence scores are methods frequently used in supporting human-AI decision-making \cite{Zhang2020,Tian2026}.

\subsubsection{Interactive Visualisation for What-if Analysis}

\begin{figure}
    \centering
    \includegraphics[width=0.95\linewidth]{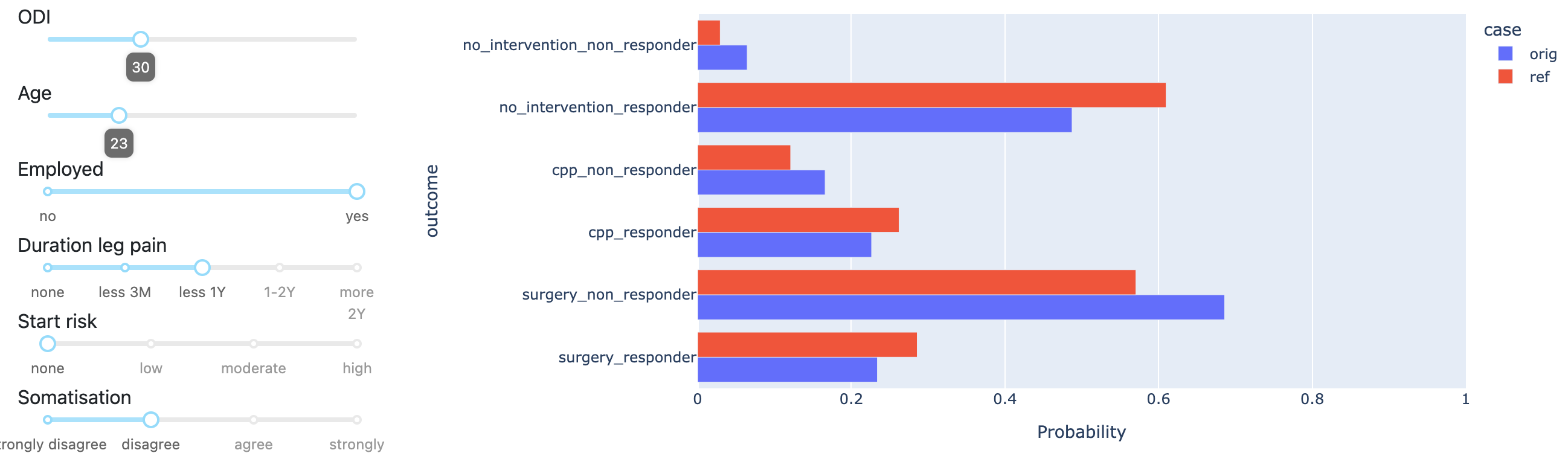}
    \caption{An interactive visualisation that allows clinicians to manually create hypothetical what-if scenarios by modifying input data using sliders. The predictions based on the initial patient information are shown in blue in the bottom bar (\textit{orig}), while the predictions based on the slider values are shown in red in the top bar (\textit{ref}).}
    \label{fig:sliders}
    \Description{The image depicts an interactive element for what-if analysis. On the left are sliders, which allow the modification of patient information. On the right, the DSS predictions are shown based on the patient information. For each treatment outcome, two bars are shown: one based on the original patient information, and one based on the patient information adjusted using the sliders. This allows the comparison of changes in outcomes based on different input factors.}
\end{figure}

In addition to the questions, we implemented an interactive element that allows decision-makers to change the patient information, i.e., input data, using sliders, with the corresponding updated predictions displayed immediately (Fig.~\ref{fig:sliders}). The updated and original predictions for each treatment option are presented in two stacked bar charts, allowing for comparison between them. Creating what-if scenarios is a common method for examining the performance of machine learning models, which is typically intended for developers \cite{Wexler2019}. As such, decision-makers ``can perform counterfactual reasoning, investigate decision boundaries, and explore how general changes to data points affect predictions, simulating different realities'' \cite{Wexler2019}. 
Similar to the question type, which displays hypothetical questions by generating perturbations computationally, this interactive visualisation allows the manual creation of input perturbations or what-if questions.

\section{Methods}
\label{sect:methods}

We conducted individual, qualitative semi-structured interviews to gain initial insights into the perceived usefulness of a DSS that provides questions alongside predictions. The interviews are part of a larger study on human-AI decision-making in the clinical context. In addition to collecting feedback on the prototype, we investigated how clinicians work with the existing DSS to make decisions in clinical practice. The general decision-making process with the DSS is, however, the subject of another paper, for which we collected additional data via an online survey and from a second medical cohort in which clinicians use a different DSS.
In this paper, we only analyse the interviews with a focus on the prototype.
We received approval from the ethics committee of our institution.

\subsection{Participants}

Our participants are clinicians who work in the same spinal department and use a decision support system to determine treatment for chronic low back pain, as described in section~\ref{sect:use-case}. We were able to recruit seven of the eleven clinicians (63.6\%) working with the DSS in clinical practice. The DSS is not used in other hospitals, thus the small population size. A researcher affiliated with the hospital who is also an author of this paper introduced us to a clinician, who then provided us with contact details of their colleagues.
Among the participants were five orthopaedic surgeons, one physician, and one physician assistant. Although the level of professional experience in orthopaedic practice varies among our participants, most of them are quite experienced; six participants have at least six years of professional experience. Moreover, each participant has at least three years of experience using the DSS in clinical practice, with five participants having six or more years of experience using the DSS. All participants use the DSS regularly, with the majority using it for patient consultation and to inform decision-making regarding treatment. The characteristics of our participants are summarised in Table~\ref{table:participants}.
Each participant was informed about the purpose of this research and how their data would be used and stored, participated voluntarily and gave their written consent. They were compensated with 15€/h for participating in the interview.

\begin{table}[t]
    \caption{Characteristics of our participants. While some participants use the DSS for both triage and consultation, the DSS Usage column indicates the primary use. Frequency of use was measured on a six-item scale ranging from ``Once or twice a month'' to ``Several times a day''.}
    \label{table:participants}
    \centering
    \begin{tabulary}{\linewidth}{Lp{3cm}LLp{2cm}p{3.1cm}}
    \toprule
         \textbf{ID} &
         \textbf{Profession} &
         \textbf{Prof. Experience (years)} &
         \textbf{DSS Use (years)} &
         \textbf{DSS Use} &
         \textbf{Frequency of DSS Use}
         \\
    \midrule 
         P1 &
         Orthopaedic Surgeon &
         9+ &
         6-8 &
         Consultation &
         Several times a week
         \\
         P2 &
         Orthopaedic Surgeon &
         3-5 &
         3-5 &
         Triage &
         Several times a week
         \\
         P3 &
         Physician Assistant &
         6-8 &
         3-5 &
         Triage &
         Several times a week
         \\
         P4 &
         Orthopaedic Surgeon &
         9+ &
         9+ &
         Consultation &
         Several times a month
         \\
         P5 &
         Orthopaedic Surgeon &
         9+ &
         6-8 &
         Consultation &
         Once or twice a week
         \\
         P6 &
         Physician &
         9+ &
         9+ &
         Consultation &
         Several times a week
         \\
         P7 &
         Orthopaedic Surgeon &
         6-8 &
         6-8 &
         Consultation &
         Several times a week
         \\
    \bottomrule
    \end{tabulary}
\end{table}

\subsection{Expert Interviews}

We conducted semi-structured interviews. All interviews took place in person at the practitioners' offices and were conducted by the first author. With an introduction to the interview process and the collection of the consent forms, five interviews lasted about 60 minutes, while two lasted about 30 minutes due to the limited availability of the two participants. 
The general procedure of the interviews consists of the following three parts.

In the first part, we asked the participants some general questions regarding their profession, when they started using the DSS, the frequency of use, and whether they perceived it as useful. We then showed them our prototype displaying treatment predictions based on a randomly selected patient case from our synthetic data set. We asked whether the visual representation looked familiar to them and whether the patient case was realistic. Some participants then guided us through their reasoning for that patient case, explaining how they made sense of the patient information and what treatment they would recommend.

In the second part, we examined the perceived usefulness of the questions and what-if scenarios presented by our prototype (Table~\ref{table:questions}). The question types remained the same for all randomly selected patient cases.
We showed the clinicians the questions one by one and, finally, the interactive what-if visualisation. 
For each question, we asked them how they perceived it: whether they found it useful, whether they saw any value in it, whether it would help them in their decision-making, or whether it would annoy them. 

In the third part and at the end of the interview, we asked if they found a particular question helpful, whether they had anything to add, and explained the purpose of our approach of a system that outputs questions, i.e., to stimulate reflection during decision-making and mitigate overreliance.

We collected and analysed data in a cyclical way. Each interview was analysed before the next interview was conducted. Hence, previous interviews informed subsequent interviews, while the general procedure remained the same. This approach allowed us to compare findings and to seek clarification in further interviews. Besides, after the first interview, two authors discussed the procedure and slightly adjusted it to ensure that certain aspects were covered during the interviews, such as understanding the decision-making workflow and ensuring that participants understand the purpose of the questions in our prototype.

\subsection{Data Analysis}

We audio-recorded the interviews and transcribed them using noScribe\footnote{\url{https://github.com/kaixxx/noScribe}. Version used: 0.6.2}. We checked the transcripts for correctness by going through the generated transcripts while listening to the recordings. We edited the transcripts for readability and removed personally identifiable information to preserve the anonymity of our participants.

To analyse our data, we used thematic analysis, an open and flexible coding procedure \cite{Braun2006}.
The same author who conducted the interviews also analysed the data, hence they were already familiar with the data.
After each interview, we refined the codebook by combining similar codes, creating relations, and updating (preliminary) themes. 
We identified codes and themes primarily inductively (i.e., data-driven), meaning we did not use any pre-existing framework or codebook \cite{Braun2006}. Our research question nevertheless guided us in a deductive manner to understand how clinicians perceive our prototype.
Furthermore, our analysis remained at the semantic level, meaning we focused on what the participants said without interpreting any underlying meanings that might have been constructed, for example, by the hospital's organisational structures \cite{Braun2006}.

\section{Qualitative Results}
\label{sect:findings}

We presented the participants with our prototype (Fig.~\ref{fig:prototype}) that displays a patient case, the treatment prognoses, seven different questions about the patient case (Table~\ref{table:questions}), and an interactive visualisation for what-if analysis (Fig.~\ref{fig:sliders}). All participants confirmed that our prototype looked familiar to them, as the way predictions were displayed resembled that of the DSS they use in clinical practice. 
The participants also confirmed that our generated cases represent realistic cases that they could encounter in their work. As P2 explained, ``We see all kinds of patients. So I don't think there's anything unrealistic over here''.

In general, the clinicians we spoke to rated the DSS they use in clinical practice as helpful for their decision-making. As P4 noted, ``we try to look for resources to help us with that decision, and the DSS is one of them''. Nevertheless, four participants mentioned that the biggest limitation is that the DSS does not take into account other diagnostics, such as MRI scans. P1 stated, ``if you can machine learn the MRI and have the results introduced in this decision-making tool as well, then we could take a next step''. For this reason, the participants also pointed out that they do not regard overreliance on the predictions as a problem for themselves, as the DSS is just one source of information among many. Besides, given that the participants have been using the DSS for several years, they considered their understanding of how it works to be good and thus also developed a `feel' for when they could trust the DSS predictions and when they could not.

Below, we present the seven themes we identified with regard to our research question, as well as recommendations based on feedback from the participants.

\subsection{T1: Questions Are Not Particularly Helpful in Stimulating Reflective Thinking Among Experienced
Decision-Makers}
\label{sect:theme-one}

Our participants, who have several years of experience working with the DSS in practice, perceived the questions of our prototype as not too helpful for their decision-making. As P4 admitted, ``I don't really see the relevance of asking questions to the physician''.
This is mainly for two reasons.

First, \textit{the questions concerned aspects that our participants take into consideration anyway}, such as the location of previous surgery. As P2 explained, ``the problem is that I think this is quite a difficult hospital to start with because we see terribly complex patients with lower back problems for over 40 years who have seen nine to 10 specialists previously. So we're quite well trained in making a decision about who would benefit from an operation, and who would definitely not''. P1 demonstrated this expertise by providing a detailed rationale of their decision-making process based on our synthetic patient case. 

Second, \textit{the questions did not address the decision-making context appropriately}. P2 mentioned that the question addressing age is not relevant because ``I think 90\% of our patients is 50 plus''. Moreover, P3 added that age is not the most crucial factor to consider: ``It's [age] not as important as ODI, back and leg pain, but on the side, it's important''.
Similarly, P1 rated the question about the confidence score of the DSS as inapplicable, as this did not fit the clinical reasoning process, ``whether this is with a 45\% confidence interval, this is not how we work in the clinic'' (P1). P6 confirmed this by saying, ``We never believe that it [DSS] is 100\% confident. I don't think this will be very helpful''. 

Nevertheless, participants pointed out several benefits, as we will discuss below. In addition, we will mention aspects that should be considered when formulating questions or designing similar friction strategies to encourage reflection.

\subsection{T2: Questions Can Support Decision-Making by Serving as Reminders to Consider Information} 
\label{sect:theme-two}

Four participants (P2, P3, P6 \& P7) pointed out that questions are useful in serving as reminders to consider information. 
P2 noted that after a busy day, when attention is low, they might overlook (obvious) factors. As they explained, ``In the reference letter, they [previous doctors] quite often refer to an MRI scan, which isn't included. If it isn't included and I'm at the end of the day, I'm really dreadful and I need a rest, then I think I need to see the surgery because it's really clearly stated that there is no compression on the MRI scan. If it's early in the morning, then I send an order to the secretary, please ask for the MRI scan so we can have a second look''. In this scenario, a question like ``Did you look at the MRI scan?'' (P2) could ensure that certain factors are considered, acting as a form of safeguard. In addition, P1 mentioned that they rarely consider age, ``the age is not really on my mind when I use the [DSS]''. This is precisely where questions could be helpful, by reminding clinicians to consider certain facts. P6 corroborated this, ``In the extreme cases, like below 20 and above 70, then it's good to get a reminder, like, did you check the age''. P5 also confirmed that questions could help them pause for a moment and think things over again.

\subsection{T3: Modifying Data Can Help to Make More Accurate Predictions}
\label{sect:theme-three}

We showed the participants an interactive visualisation that allows them to manually create what-if scenarios by adjusting the input data using sliders (Fig.~\ref{fig:sliders}). Before showing the interactive plot to one participant, they expressed the need for these what-if scenarios themselves, ``if you could see a bar here, that is not as bright as this one, that would sort of predict what [the outcome] would be if there were no prior surgeries, I think that would help'' (P4). 
All participants were pleasantly surprised to see the immediate change in the predictions, even though, in theory, they know about the effects of certain variables. As P4 further noted, ``from the literature, we know it's [the impact of certain factors] true''. 

The current DSS used in clinical practice does not allow for the adjustment of information. The participants, however, confirmed that changing information from the patient questionnaire can be helpful for various reasons. First, factors that are not relevant could be removed, such as previous surgeries at a different location on the spine. P5 commented on our prototype, ``if I could just delete the previous surgery, what is the DSS saying then? This is really helpful''. 
Second, patients' symptoms or circumstances may change between the time they complete the questionnaire and the day of their consultation, which may take place months later. P6 explained, ``sometimes BMI is a problem and by the time of consultation, not any more, because we have quite long waiting times. Or sometimes they stopped smoking''.
Third, it happens that patients misunderstand the questionnaire and `flip' the pain score, for example. P7 mentioned, ``sometimes I have patients that fill in a 10 [highest pain score] but they actually mean it's perfect or they fill in a 0 [lowest pain score] because they say it doesn't work at all. So they read it but they misinterpreted the question''. Fourth, patients might exaggerate their symptoms in the hope of being treated more quickly. ``A lot of people think that, if they exaggerate a little bit on the score forms, it could be beneficial for them'' (P5).

\subsection{T4: What-if Scenarios Can Help to Consider Alternative Courses of Action} 

The hypothetical question that is based on counterfactual reasoning (question seven, e.g., \textit{``Is it possible to reduce BMI? This would increase the predicted effectiveness of surgey by 15\%''}), allows clinicians to consider a different course of action. In doing so, another treatment could be recommended before going with the one the DSS recommends, such as surgery. P2 explained, ``When we see an increase of 30\% chance of succeeding by treating the somatisation, then we could send [the patient] to the psychiatrist, and then we re-evaluate, and make a decision whether to operate or not''. Consequently, this forward-looking hypothetical question, which asks about the possibility of changing the factor that has greatest influence on the prediction, could help clinicians consider alternative treatment options or pre-treatment.

\subsection{T5: What-if Scenarios Can Inform Discussions With Patients}

Both the hypothetical question based on counterfactual reasoning and the interactive what-if visualisation were recognised as helpful in persuading patients to change their lifestyle during consultation. P3 explained, ``If you know, okay, if the patient stops smoking and drops five kilograms the chances are really increasing by 25\%. That would be really handy, then you can talk to the patient''. 
Similarly, P1 mentioned that the hypothetical question could be helpful during treatment, ``suppose you end up recommending the patient to conservative care, and they go to this cognitive pain management program. Then, over there they should talk to the patient and say, okay, if you are willing to change, if you're willing to do this, and if you expect improvement in your functional status, so maybe modify your expectations of recovery, in the end, you may have a better result''. 
Accordingly, what-if scenarios can give healthcare professionals a better understanding of the likely effects of a change in a patient's functional status. Importantly, as P6 noted, this insight can also help to persuade patients, ``we are really helped by tools that are convincing the patient to change [their] lifestyle''.

\subsection{T6: Questions and What-if Scenarios Can Help to Be More Critical of the DSS Predictions}
\label{sect:theme-four}

Participants considered it instructive to see the impact of a feature on the prediction.
At the beginning of the interview, P4 pointed out that it would be helpful for them to see the most contributing features of a prediction, ``if you just get in a box next to it, the three main drivers for the prediction, I think that could really help''. For this reason, the interactive what-if visualisation was found to be insightful. P5 mentioned, ``I think what you just showed me, that you can see how important all the findings are on the prediction, that would help''. P7 corroborated the value of the what-if scenarios, ``Yeah, because now you at least can see which factors decided to steer it [the prediction] in a certain direction''. 
For the same reason, the hypothetical question based on counterfactual reasoning was also perceived as insightful. Part of the question provided information on the estimated change in effect, e.g., the chances of the surgery being successful increase by 25\% if the patient changes their expectations regarding recovery. When seeing the indicated change in outcome, P4 exclaimed, ``Wow, that's impressive''. P5 explained that seeing this change is helpful in understanding the DSS, ``it gives more insight, of course, to how the whole tool works''.

When showing the participants the question regarding age, i.e., whether age is relevant to consider when a patient is close to a certain threshold, participants were wondering about the built-in threshold of 50 years. It is important to note that the DSS, which the clinicians have been working with for several years and on which our prototype is based on, uses the same age threshold.
P2 questioned the adequacy of the threshold value, ``I understand there is a threshold for age, but I don't know whether it should be 50'' (P2). P5 raised similar concerns, ``The only thing I think is that 50 years is a strange cut off moment''. Consequently, the question of our prototype helped them to become more aware of how the DSS works and, ultimately, more critical of it. 

\subsection{T7: Questions and the What-if Scenarios Could Be Helpful for Novice Clinicians}
\label{sect:theme-five}

Our participants, who have multiple years of experience, noted that less experienced colleagues could potentially benefit from questions as well as from the interactive what-if visualisation.
Accordingly, P7 highlighted the benefit of questions serving as reminders for novice clinicians, ``If they've missed it, because they didn't ask it [the patient during consultation]. And so, uh, if it would come as a pop-up, did you ask about the previous operations, then it can help them''. Similarly, P4 noted ``I think if less experienced people don't really know how it [DSS] works, for instance, with the previous surgery, they might say no to a patient for surgery, which actually might be suitable for surgery''. In view of this, P2 explained that the interactive visualisation could help clinicians become familiar with the DSS, ``I think it would be really helpful for new colleagues or colleagues who start using the [DSS]. And just learning how to read it, how to use it. Don't be quite blind and only use the DSS because when you see this really small change [...] it makes a huge difference in the response [prediction]''. Participants therefore regarded our prototype as a potential training tool that could help inexperienced decision-makers use the DSS and scrutinise its predictions more critically.

\subsection{Recommendations for Designing Friction Strategies}
\label{sect:theme-six}

Based on our participants' feedback, we identified five aspects that should be considered when designing friction strategies. 

\textbf{Friction strategies should encourage unbiased reasoning.}
As P7 explained, ``So you're always biased, also in the way you're trained. And so a surgeon will always think in surgical solutions''. 
It can thus be beneficial to promote divergent thinking, for example through counterfactual questions, that help decision-makers form a `second opinion' that deviates from entrenched thought patterns. In doing so, the larger solution space can be investigated.

\textbf{Friction strategies should account for the varying expertise and individual differences of decision-makers.}
Some of our questions were perceived differently by our participants. For the question about the confidence of the decision-maker, for example, P6 stated that this question would not help them, whereas P5 mentioned, ``This could be a nice question. Because I think this is the problem, for me, at least''. In addition, P6 pointed out that the trigger for a question about age is likely to be different for physicians and surgeons, ``for me, it's [age threshold] below 20 and above 70. And I can imagine that for the surgeons the range when they want to get an alert is different, like, maybe above 60''.
Consequently, questions should ideally be personalised, particularly in view of differing decision-making requirements. 

\textbf{Friction strategies should provide actionable insights.}
The usefulness of questions in stimulating reflective thinking depends on the availability of the necessary information. P2 mentioned that although previous surgeries are indicated, ``in 90\% of cases they [other specialists] don't even mention what level [the patient] was operated on''. At the same time, patients often do not know the details about previous surgeries, as P3 explained, ``They say, I've been operated in the back. With screws? Oh, I don't know''. In view of this, P6 commented on the question about previous surgeries, ``if we know, it's helpful, but if we don't know, yeah it doesn't help you''.
Moreover, although participants preferred the question regarding hypotheticals and the interactive what-if scenarios, they also raised concerns regarding the applicability, ``But there's a lot of things the patient cannot do anything about it. Like the duration of leg pain. Or previous surgery. It's a fact'' (P3). Similarly, P2 remarked, ``I think that would be mainly helpful with things patient could change. So, his smoking behaviour, his BMI''. 
So while friction-based approaches can provide insights into the workings of the DSS (epistemic value), in certain cases clinicians may prefer a more practical approach (pragmatic value), in which the information provided must be actionable and accessible (e.g., factors that the patient can change).

\textbf{Friction strategies should be easy to understand, especially with regard to time constraints.}
As P1 explained, ``you have 20 minutes to see a patient'', and the time before consultation in which clinicians review the patient's history and previous diagnostics is limited as well. As a consequence, as P5 pointed out, ``during your outpatient clinic, you're always little bit in a hurry. Because you don't have that much time''. P2 explained that our prototype would not be too helpful for them in their task of triaging patients, ``because I have no time to play with all those graphs''. Another participant noted that the visualisation was easier to understand and thus likely to be more efficient, ``If you look at this, with, like, different colours and figures, it is better for surgeons. But if you have to read a lot, it's not what we're good at'' (P5). Frictional strategies should therefore allow for (productive) interruptions in which reflection can take place while still fitting into existing workflows without being too complicated. 

\textbf{Friction strategies should fit the decision-making context and workflow.} 
The participants mentioned that there are various decision-making stages throughout the patient trajectory. As such, for the question regarding the assumptions of the decision-maker, P3 explained that the question could be helpful after seeing the patient, ``If you see it [the DSS] after you saw the patients, then you can check it [one's assumptions] a bit. So then you can say, does the prediction match my assumption?''. At the same time, however, P1 noted that they already have certain assumptions before seeing the patient, ``if you have this decision tool and you know this already before you see the patient, to some extent, you are a little bit biased when you see the patient''. 
Questions or other friction strategies could therefore intervene at different stages in the decision-making process, and accordingly address different information, such as helping to overcome anchoring bias before the consultation or to reflect on patient preferences after the consultation. This also hints at the multifaceted decision-making process in clinical context. As P7 explained when seeing a generated patient case, ``there's no clinical story here. And so, it's, it's only the model''.

\section{Discussion}
\label{sect:discussion}

Based on a real-world decision-making context in which clinicians use a DSS in clinical practice, we presented our qualitative results on the perceived usefulness of two friction strategies, namely questions and what-if analysis.

We found that questions did little to stimulate reflective thinking in this particular use case (T1). Our participants have extensive experience both professionally and in working with the DSS. As such, they are trained to make carefully considered decisions based on the patient information. Accordingly, participants also did not regard overreliance as a problem for them, even though they confirmed that the DSS’s predictions influence their decision-making. Importantly, as our questions were designed to encourage reflection, it is only natural that they were not perceived as particularly useful in situations where decision-makers already reflect. As such, most of our questions addressed factors that the clinicians already take into account (T1).
As highlighted by P2 (see T2), however, something that is obvious in the morning, when one's mind is still fresh, might be less obvious at the end of the day, when one is tired. Hence, a question about an `obvious' fact might be helpful in certain circumstances.
Therefore, as our participants mentioned, questions can serve as reminders to consider information (T2), particularly when attention is low. This is in line with findings from the aforementioned study on mitigating cognitive biases in ophthalmology \cite{Bach2023}. \citet{Bach2023} found that their `consider the opposite' strategy could help clinicians avoid overlooking crucial details, as one of their participants stated, ``Sometimes you get tired, so if the program said 'Something is there', and I scale up [the image] and then I think 'Yes, I didn’t notice that myself'''.

Nevertheless, the finding that questions are not helpful in stimulating reflection stands in contrast to other studies, which report positive effects of questions in that regard \cite{Danry2023,Reicherts2022,Li2026a,Tian2026}.  \citet{Reicherts2022} recruited participants who have (some) experience with the decision task, while others \cite{Danry2023,Li2026a,Tian2026} relied on laypeople recruited via Prolific or Credamo who were unfamiliar with the task. In addition, the tasks in \cite{Danry2023,Li2026a,Tian2026} are of general purpose each of which consists of a ground truth, such as assessing the correctness of socially divisive statements \cite{Danry2023}, predicting diabetes \cite{Li2026a}, and predicting house prices \cite{Tian2026}. 
In view of this, the opposing findings suggest that the effectiveness of questions in stimulating reflection depends on the task and its complexity (whether there is a ground truth or not), as well as two dimensions of the decision-maker's expertise: namely, their familiarity with the task and with the DSS. 

The assumption that expertise and task familiarity influence the effectiveness of questions is further corroborated by our participants, who emphasised the (potential) value of questions and what-if scenarios for decision-making by less experienced clinicians. Participants were of the opinion that less experienced clinicians may rely more heavily on the DSS predictions since they have less professional experience to draw from. Hence, participants also mentioned that our prototype could be helpful for novice clinicians in learning how to use the DSS (T7). Similarly, \citet{Bach2023} found that, ``All ophthalmologists expressed that the debiasing techniques could be a learning tool for inexperienced screeners to heighten their diagnostic accuracy''. The findings that friction-based approaches, such as questions and what-if scenarios, hold educational benefits are therefore consistent with previous research findings.

In view of this, friction strategies can serve different purposes, namely stimulating reflective or critical thinking (T1), serving as reminders to consider information (T2), promoting practical action (T3, T4, T5), providing insights into how the model works (T6), and acting as an educational tool (T7). 
This finding corresponds to discussions in the field of explainable AI. \citet{Langer2021} point out that explanations hold different desiderata, such as increasing the confidence of the decision-maker, which relates to T1; increasing the trustworthiness of the systems, which relates to T6; or helping decision-makers to learn how to use the system, which relates to T7. In view of this, our reflective questions and what-if scenarios pursue similar objectives as explainable AI. We therefore view questions and explanations as complementary, as also utilised by \citet{Li2026a}. %
In any case, relevant studies \cite{Tian2026,Li2026a,Danry2023,Reicherts2022} currently primarily view questions as stimulating reflection (T1) to reduce overreliance on AI or to improve decision accuracy. While certainly commendable, similarly ``good'' outcomes might be also achieved by focusing on the other aspects of questions, such as becoming familiar with the DSS (T7) or promoting action such as stimulating discussions with others (T5).

With regard to our questions, it might be that questions beyond those our prototype presented prove more effective in stimulating reflective thinking. We noticed that the clinicians we spoke to responded more positively to questions that included the importance of a factor, for example, by showing the impact of previous surgery on the prediction. This information not only gives insights into how the DSS works (T6), similar to explanations, but also provides a reason why decision-makers should entertain a particular question, and enables them to compare their own judgment with the influence of that factor. 
To extract the influence of single features, we used feature perturbations as explained in section~\ref{sect:question-generation}. To generate a different set of questions than the ones we have used, it would be conceivable to utilise feature contributions, a common approach in explainable AI. Instead of providing the feature contribution in the form of a statement, \textit{``outcome y because of factor x''}, the explanation could be reformulated as a question, e.g., \textit{``Does outcome y follow from x?''}. This approach was also used in the aforementioned study by \citet{Danry2023}. Importantly, this type of question enables decision-makers to reconsider information of the DSS, e.g., ``Is this factor the most important one?''. 

Our participants rated the hypothetical question based on counterfactuals and the interactive what-if analysis as most useful (T3, T4, T5, T6). These two methods show how changes to a specific input affect the predicted outcome. In doing so, it is possible to investigate alternative outcomes to the current DSS predictions. Similarly, according to \citet{Bach2023}, ``ophthalmologists stated that ‘consider the opposite’ would be the most helpful in ensuring that they do not miss any critical details''. 
In addition, participants in the study by \citet{Bach2023} mentioned that they know how to make decisions based on given information. 
In view of this, DSS should go beyond predicting the most likely outcome and instead allow for the consideration (and simulation) of alternative outcomes.

Overall, friction strategies should encourage open-minded thinking and help to form opinions that differ from pre-existing assumptions. As one participant mentioned, surgeons always look for surgical solutions. \citet{Danry2023} also point out that personal factors and preexisting beliefs (on those divisive topics) influence the decision-making process. 
Hence, it might be useful to tailor the questions to individual needs, such as the experience levels and informational requirements within a particular specialism, for example those of a (novice) surgeon or a (senior) physician. \citet{Reicherts2022}, who present a chatbot that scaffolds reasoning, suggest to personalise questions by configuring the chatbot based on preferences and expertise, since some of their participants also rated the questions posed by the chatbot as irrelevant to their decision-making. %
In view of the above, it is therefore crucial to generate `the right' question that fits the situation and the experience of the decision-maker. 

Given the time constraints in (clinical) decision-making, it is particularly important to balance a smooth workflow with meaningful interruptions. 
\citet{Bach2023} identified similar concerns regarding time constraints, and reported that their mitigation strategies were perceived as too intrusive, leaving participants feeling as though they were fact-checked by the system. %
Likewise, participants in the study by \citet{Reicherts2022} considered some of the questions of the chatbot as too leading and suggestive, and thus as ``too pushy''. Our questions and what-if analysis, on the other hand, were not perceived as too intrusive or pushy. Nevertheless, for a question to be meaningful and effective in a given workflow, the phrasing matters.

Finally, our friction strategies, like others, focus on the decision-maker, here, the clinician. It might be useful, however, to also address other stakeholders, such as patients, with questions or other frictional aids that help them to reflect on different aspects. 
In our use case, the DSS takes a patient-reported questionnaire as input. Patients could be assisted in completing the questionnaire to improve the reliability and quality of the data. As such, the question regarding previous surgeries (when and at which region of the spine) could be directed to patients rather than clinicians. With self-reported data, experts, i.e., clinicians, will nevertheless have to assess the reliability of the data provided. Still, frictional approaches might also be beneficial to empower other stakeholders and thereby improve the overall decision-making process.

\section{Limitations and Future Work}
\label{sect:limitations}

Our work has some limitations. Our sample size ($n=7$) is too small to make general conclusions on the usefulness of questions in human-AI decision-making. Our aim, however, was to gain insights drawn from a specific, real-world decision-making context involving experienced decision-makers. For this, we replicated a DSS used in clinical practice ($DSS_{clinic}$) and were able to recruit the majority of clinicians (7 out of 11, 63.6\%) who have been working with $DSS_{clinic}$. We believe that the ecological validity of our research is helpful in providing meaningful information for future studies, especially since our results are in line with previous studies from clinical practice and contradict the findings derived from general tasks discussed in the literature.

While we asked clinicians about the perceived usefulness of questions, we do not know the actual effects of questions on the decision-making process. It may be that our questions would have had stimulated reflection under certain circumstances (e.g., when attention is low), even though the participants in the interviews assumed that they would have considered this line of reasoning anyway.
Besides, some participants initially misunderstood the purpose of the questions provided by our prototype. As such, one assumed the questions would be addressed to patients as part of the questionnaire, while others assumed the questions were part of the interview questions. Although this may have had a minor impact on our results, it rather shows that receiving questions from a DSS is an uncommon interaction.
Furthermore, the concerns raised by our participants relate specifically to the implementation of our prototype. We implemented seven different questions, whereas it is possible to generate more and different questions. This means, obviously, that the exact implementation matters.
Future studies could therefore (quantitatively) examine the impact of questions on decision-making, especially in relation to the cognitive engagement of decision-makers, using a larger and more varied sample (e.g., in terms of experience or time of day) as well as a broader and more diverse range of questions in order to identify the `right questions' (at the right moment) that stimulate reflection.

Although we focused a real-world decision task for which we replicated a DSS, the decision-making process in clinical practice involves more factors than the information provided by the DSS. During triage, the clinicians review the patient's medical history provided by previous doctors, as well as additional diagnostic findings. 
During consultation, the patient might reveal additional contextual information that requires the clinician to reconsider previous assumptions or treatment decisions. 
The diverse sources of information and multiple decision steps highlight the complexity of clinical decision-making. For decision support systems and friction strategies to be effective, a better understanding of the decision-making process involving DSS in practice is necessary \cite{Eckhardt2026,Raees2026}.

\section{Conclusion}
\label{sect:conclusion}

In this research, we set out to investigate the potential of a prototype system that presents questions and what-if scenarios to clinicians during their decision-making process as part of a real-world decision-task. To this end, we replicated a decision support system ($DSS_{clinic}$) that has been used in clinical practice for several years to predict treatment outcomes for chronic low back pain. In addition to the predictions, our prototype outputs \textit{questions} based on the functionality of the DSS, its input data (i.e., tabular patient information), and its predictions. The prototype also allows clinicians to manually create \textit{what-if scenarios} by adjusting the input data using sliders and observing the corresponding changes in the outcomes.
We conducted semi-structured, in-situ interviews to gather feedback on our prototype from clinical spine experts ($n=7$) who work with $DSS_{clinic}$.

Using thematic analysis, we found that experienced decision-makers generally did not consider our questions to be particularly helpful in stimulating reflection, as the questions concerned factors that they already take into account (T1). Participants nevertheless mentioned that questions could serve as reminders to check certain information, especially when attention is low (T2). The interactive visualisation for what-if analysis, was rated useful, as it allows to modify input data, thereby leading to more accurate predictions (T3), it helps to consider alternative courses of action (T4), and it can inform discussions with patients (T5). In view of this, the what-if scenarios could help to improve decision-making and consequently patient care.
Moreover, clinicians preferred both the what-if scenarios and the hypothetical question based on counterfactual reasoning, as they provide insights into how the DSS works (T6). Overall, participants mentioned that our prototype could be especially useful to novice clinicians in learning how to use the DSS (T7). 
These findings highlight that questions and what-if analysis can serve different purposes.
Finally, based on the feedback from clinicians, we provided recommendations for the design of friction strategies.

Our qualitative results from an ecologically valid use case corroborate the findings from another study from clinical practice regarding the potential of friction strategies serving as memory aids and learning tools \cite{Bach2023}. At the same time, our findings contradict the positive effects of questions in stimulating reflection that have been observed in other studies \cite{Tian2026,Li2026a,Danry2023,Reicherts2022}.
It is therefore important to evaluate friction approaches in real-world settings.

\section*{Generative AI Usage Statement}
The authors used DeepL Translator to adjust individual sentence structures by translating the English sentence into another language and then back into English. Grammarly (with AI features turned off) was used to polish some sentences. Moreover, we used noScribe (version 0.6.2) to transcribe the audio recordings of our interviews, from which we have used quotes throughout the article. We have checked the transcripts for accuracy, as described in the methods section. Apart from that, no other generative AI tool was used in the creation of this paper.

\begin{acks}
\end{acks}

\bibliographystyle{ACM-Reference-Format}
\bibliography{main}

\appendix

\end{document}